\def\rfr#1{eq. (\ref{#1})}

\def\leti{Lense--Thirring}
\def\Rfr#1{Eq. (\ref{#1})}

\def\derp#1#2{\rp{\partial{#1}}{\partial{#2}}}
\def\bar{\begin{eqnarray}}
\def\ear{\end{eqnarray}}

\def\eqi{\begin{equation}}
\def\eqf{\end{equation}}
\def\eqia{\begin{eqnarray}}
\def\eqfa{\end{eqnarray}}
\def\rp#1#2{{#1\over#2}}
\def\ct#1{\cite{#1}}
\def\lb#1{\label{#1}}

\def\oc2{$\mathcal{O}(c^{-2})$}

\documentclass[11pt]{article}
\usepackage{amsmath,amsthm,amscd,amssymb}
\usepackage{latexsym}
\usepackage{graphicx,epsfig}

\begin{document}

\noindent{\bf \LARGE{Some comments on the recent results about the
measurement of the Lense-Thirring effect in the gravitational
field of the Earth with the LAGEOS and LAGEOS II satellites.}}
\\
\\
\\
{Lorenzo Iorio}\\
{\it Dipartimento Interateneo di Fisica dell' Universit${\rm
\grave{a}}$ di Bari
\\Via Amendola 173, 70126\\Bari, Italy
\\e-mail: lorenzo.iorio@libero.it}

\begin{abstract}
In a recently published paper Ciufolini reports on the so far
performed tests aimed at the detection of the general relativistic
gravitomagnetic Lense-Thirring effect in the gravitational field
of the Earth by means of the analysis of the laser-ranged data of
the existing LAGEOS and LAGEOS II geodetic satellites. In this
paper we will critically discuss his claims by showing that the
total error, mainly due to the systematic bias due to the
mismodelling in the static and time-varying parts of the
multipolar expansion of the Newtonian terrestrial gravitational
potential, is larger than that claimed by Ciufolini. E.g., the
systematic error due to the mismodelling in the static part of the
geopotential in the tests performed with the EGM96 Earth gravity
model and the combination involving the nodes of LAGEOS and LAGEOS
II and the perigee of LAGEOS II realistically amounts to more than
80$\%$ (1-$\sigma$): the claimed total uncertainty, including also
the non-gravitational perturbations which especially affect the
perigee of LAGEOS II, is, instead, 20-25$\%$. The claimed accuracy
in the more precise tests performed with the 2nd generation
CHAMP-only EIGEN2 Earth gravity model and a combination involving
the nodes of LAGEOS and LAGEOS II over 10 years is 18$\%$. With
numerical simulations we will show that, instead, it is $\leq
51\%$ (1-$\sigma$) if the impact of the secular variations of the
even zonal harmonics over a so long observational time span ($\sim
14\%$) is accounted for.
\end{abstract}
\newpage
\tableofcontents
\newpage
\section{Introduction}
Recent years have seen increasing efforts aimed to directly
detecting various phenomena connected to the general relativistic
gravitomagnetic field \ct{mash01, ciu95, tart, scha} of the
rotating Earth. It should be noted that, according to K. Nordtvedt
\ct{nor03}, the multidecadal analysis of the Moon'orbit by means
of the Lunar Laser Ranging (LLR) technique yields a comprehensive
test of the various parts of order $\mathcal{O}(c^{-2})$ of the
post-Newtonian equation of motion. The existence of
gravitomagnetism as predicted by the Einstein's General Theory of
Relativity would, then, be indirectly inferred from the high
accuracy of the lunar orbital reconstruction. In \ct{norlag} the
same arguments are applied to the radial motion of the LAGEOS
satellite.

The extraordinarily sophisticated and expensive Gravity Probe B
(GP-B) mission \ct{eve, evetal01} has been launched in April 2004;
it is aimed at the detection of the gravitomagnetic precession of
the spins \ct{sch60} of four superconducting gyroscopes carried
onboard at a claimed accuracy of 1$\%$ or better.

The Lense-Thirring effect on the orbital motion of a test particle
\ct{lenti} could be measured by analyzing the orbital data of
certain Earth artificial satellites with the Satellite Laser
Ranging (SLR) technique \ct{NC02}.  Up to now, the only performed
tests are due to Ciufolini and coworkers.

In this paper we will analyze the latest results presented in
\ct{ciucaz04} from a critical point of view in order to show that
the claimed accuracies are optimistic.
\section[The Lense-Thirring effect on the orbit of a test particle]{The Lense-Thirring effect on the orbit of a test particle and the strategy to measure it}
The gravitomagnetic field of a spinning mass of proper angular
momentum $J$ induces tiny secular precessions on the longitude of
the ascending node $\Omega$ and the argument of
pericentre\footnote{In their original paper Lense and Thirring use
the longitude of pericentre $\varpi=\Omega+\omega$. } $\omega$ of
a test particle \ct{lenti, ash, ciu95, ior01}
\eqi\dot\Omega_{\rm LT}=\rp{2GJ}{c^2 a^3 (1-e^2)^{3/2}},\
\dot\omega_{\rm LT}=-\rp{6GJ\cos i}{c^2 a^3
(1-e^2)^{3/2}},\lb{letprec}\eqf
where $G$ is the Newtonian constant of gravitation, $c$ is the
speed of light in vacuum, $a,e$ and $i$ are the semimajor axis,
the eccentricity and the inclination, respectively, of the test
particle's orbit.
In the terrestrial space environment the gravitomagnetic
precessions are very small: for the geodetic SLR LAGEOS
satellites, whose orbital parameters are listed in Table
\ref{para}, they amount to a few tens of milliarcseconds per year
(mas yr$^{-1}$ in the following)
{\small\begin{table}\caption{Orbital parameters ($a$ semimajor
axis, $e$ eccentricity, $i$ inclination) of the existing LAGEOS
and LAGEOS II satellites and their Lense-Thirring node precessions
$\dot\Omega_{\rm LT}$ in mas yr$^{-1}$. }\label{para}

\begin{tabular}{lllll}
\noalign{\hrule height 1.5pt}

Satellite & $a$ km) & $e$  & $i^{\circ}$  & $\dot\Omega_{\rm LT}$ (mas yr$^{-1}$)  \\

\hline

LAGEOS    &  12270    & 0.0045 &  110 & 31\\
LAGEOS II &  12163    & 0.0135 & 52.64 & 31.5\\

\noalign{\hrule height 1.5pt}
\end{tabular}

\end{table}}

The extraction of the Lense--Thirring precessions from the orbit
data analysis is very difficult due to a host of competing
classical orbital perturbations of gravitational \ct{iortid01,
iorcelmec03, iorpav01, pavior02} and non-gravitational \ct{ves99,
luc01, luc02, luc03, luc04, lucetal04} origin which have various
temporal signatures and are often quite larger than the
relativistic signal of interest. The most insidious ones are the
perturbations which have the same temporal signature of the
Lense-Thirring precessions\footnote{Also the perturbations which
grow quadratically in time are, of course, very dangerous. Those
induced by the secular variations of the even zonal harmonics of
the Earth's geopotential fall in this category, as we will see in
detail in Section \ref{zonrat}. Time-dependent periodic
perturbations with periods longer than the observational time span
may also be insidious because they would resemble superimposed
linear trends \ct{iortid01}.}, i.e. secular trends. Indeed,
whatever the length of the adopted observational time span $T_{\rm
obs}$ is, they cannot be fitted and removed from the time series
without removing the relativistic signal as well. Then, it is of
the utmost importance to assess as more accurately and reliably as
possible their aliasing impact on the measurement of the
Lense-Thirring effect.

It turns out that the perigees of the LAGEOS-like satellites are
severely  affected by the non-gravitational perturbations,
contrary to the nodes. Moreover, since the non--conservative
forces depend on the structure, the shape and the rotational
status of the satellite their correct modelling is not a trivial
task and, as we will see later, introduces large uncertainties in
the correct assessment of the error budget in some of the
performed gravitomagnetic tests.
\subsection{The gravitational error}
The even ($\ell=2,4,6...$) zonal ($m=0$) harmonic coefficients
$J_{\ell}$ of the multipolar expansion of the Earth's
gravitational potential, called geopotential, induce secular
precessions\footnote{Also the subtle non--gravitational
Yarkovsky-Rubincam force, which is due to the interaction of the
Earth's electromagnetic IR radiation with the physical structure
of the LAGEOS satellites, induces secular effects on their nodes
and perigees \ct{luc02}.} on the node and the perigee of any
near-Earth artificial satellite \ct{kau66} which, of course,
depend only on its orbital configuration and are independent of
its physical structure. Such aliasing effects are many orders of
magnitude larger than the Lense-Thirring precessions; the
precision with which the even zonal harmonics are known in the
currently available Earth gravity models \ct{jgm3, egm96, grim5c1,
eigen1s, eigen2, eigen3p, eigen-grace02s, ggm01, cg01c} would
yield errors amounting to a significant fraction of the
Lense-Thirring precessions or even larger.

Even more dangerous are the perturbations induced by the secular
variations of the low degree even zonal harmonics $\dot
J_{\ell},\ell=2,4,6$ \ct{eanes, cox}. Indeed, such perturbations
grow quadratically in time if the shifts in mas are considered and
linearly in time if the rates in mas yr$^{-1}$ are considered.
Their impact on the orbital elements of the LAGEOS satellites have
been worked out in \ct{iormor04}. It turns out that, by using the
results of \ct{eanes}, the errors induced by $\dot J_2$ would
amount to 8$\%$, 14$\%$ and $5.4\%$ for the nodes of LAGEOS and
LAGEOS II and the perigee of LAGEOS II, respectively, over an
observational time span $T_{\rm obs}$ of just one year at
1$-\sigma$ level. This clearly shows that it would be impossible
to analyze single orbital elements.

The time-dependent periodic perturbations \ct{iortid01, iorpav01,
pavior02} are less dangerous because if their periods are shorter
than the adopted observational time span they can be fitted and
removed from the time series. The most insidious tidal
perturbation is that induced by the even zonal constituent which
has a period of 18.6 years and whose nominal impact on the orbital
elements of the LAGEOS satellites amounts to thousands of mas
\ct{iortid01}. However, it turns out that it does not affect the
observables which have been adopted for the performed
Lense-Thirring tests because its main component is of degree
$\ell=2$ and order $m=0$.

\subsubsection{The linear combination approach}
The problem of reducing the impact of the mismodeling in the even
zonal harmonics of the geopotential with the currently existing
satellites can be coped in the following way \ct{ciu96}.

Let us suppose we have at our disposal N (N$>1$) time series of
the residuals of those Keplerian orbital elements which are
affected by the geopotential with secular precessions, i.e. the
node and the perigee: let them be $\psi^{\rm A},$ A=LAGEOS, LAGEOS
II, etc. Let us write explicitly down the expressions of the
observed residuals of the rates of those elements
$\delta\dot\psi^{\rm A}_{\rm obs}$ in terms of the Lense-Thirring
effect $\dot\psi_{\rm LT}^{\rm A}$, of N-1 mismodelled classical
secular precessions $\dot\psi_{.\ell}^{\rm A}\delta J_{\ell}$
induced by those even zonal harmonics whose impact on the
measurement of the gravitomagnetic effect is to be reduced and of
the remaining mismodelled phenomena $\Delta$ which affect the
chosen orbital element \eqi\delta\dot\psi_{\rm obs}^{\rm
A}=\dot\psi_{\rm LT}^{\rm A}\mu_{\rm LT }+\underset{{\rm N-1\
terms }}{\sum}\dot\psi_{.\ell}^{\rm A }\delta J_{\ell}+\Delta^{\rm
A},\ \underset{{\rm N}}{\underbrace{{\rm A=LAGEOS,\ LAGEOS\
II,...}}} \lb{equaz}\eqf The parameter\footnote{It can be
expressed in terms of the PPN $\gamma$ parameter \ct{wil93} as
$\mu_{\rm LT}=(1+\gamma)/2$.} $\mu_{\rm LT}$ is equal to 1 in the
General Theory of Relativity and 0 in Newtonian mechanics. The
coefficients $\dot\psi_{.\ell}^{\rm A}$ are defined as
\eqi\dot\psi_{.\ell}=\derp{{\dot\psi}_{\rm class}}{J_{\ell}}\eqf
and have been explicitly worked out for the node and the perigee
up to degree $\ell=20$ in \ct{NC02, iorcelmec03}; they depend on
some physical parameters of the central mass ($GM$ and the mean
equatorial radius $R$) and on the satellite's semimajor axis $a$,
the eccentricity $e$ and the inclination $i$. We can think about
\rfr{equaz} as an algebraic nonhomogeneuous linear system of N
equations in N unknowns which are $\mu_{\rm LT}$ and the N-1
$\delta J_{\ell}$: solving it with respect to $\mu_{\rm LT}$
allows to obtain a linear combination of orbital residuals which
is independent of the chosen N-1 even zonal harmonics. In general,
the orbital elements employed are the nodes and the perigees and
the even zonal harmonics cancelled are the first N-1 low-degree
ones.

This approach is, in principle, very efficient in reducing the
impact of the systematic error of gravitational origin because all
the classical precessions induced by the static and time-dependent
parts of the chosen N-1 $J_{\ell}$ do not affect the combination
for the Lense-Thirring effect. Moreover, it is flexible because it
can be applied to all satellites independently of their orbital
configuration, contrary to the butterfly configuration in which
the cancellation of the even zonal harmonics can be achieved only
for supplementary orbital planes and identical orbital parameters.
Apart from the first orbital element which enters the combination
with 1, the other elements are weighted by multiplicative
coefficients $c_i(a,e,i)\neq 1$ which are built up with
$\dot\psi_{.\ell}$ and, then, depend on the orbital elements of
the considered satellites. Their magnitude is very important with
respect to the non-gravitational perturbations, which in general
are not cancelled out by the outlined method, and to the other
time-dependent perturbations of gravitational origin with
$\ell\neq 2,4,6,..,m\neq 0$. Values smaller than 1 for the $c_i$
coefficients are, in general, preferable because they reduce the
impact of such uncancelled perturbations. It is important to note
that the order with which the orbital elements enter the
combination is important: indeed, while the systematic error due
to the even zonal harmonics of the geopotential remains unchanged
if the orbital elements of a combination are exchanged, the
coefficients $c_i$ do change and, consequently, also the
non-gravitational error. The best results are obtained by choosing
the highest altitude satellite as first one and by inserting the
other satellites in order of decreasing altitudes.

This method was explicitly adopted for the first time in
\ct{ciu96} with the nodes of the LAGEOS satellites and the perigee
of LAGEOS II. The obtained combination is
\eqi\delta\dot\Omega^{\rm LAGEOS }_{\rm obs
}+c_1\delta\dot\Omega^{\rm LAGEOS\ II}_{\rm obs
}+c_2\delta\dot\omega^{\rm LAGEOS\ II}_{\rm obs }\sim \mu_{\rm
LT}60.2,\lb{ciufform}\eqf where $c_1=0.304$, $c_2=-0.350$ and 60.2
is the slope, in mas yr$^{-1}$, of the expected gravitomagnetic
linear trend. \Rfr{ciufform} is insensitive to the first two even
zonal harmonics $J_2$ and $J_4$. It has been used in
\ct{science98} when the level of accuracy of the JGM3 \ct{jgm3}
and EGM96 \ct{egm96} Earth gravity models, available at that time,
made it necessary to consider a combination of observables which
is independent of errors in both $J_2$ and $J_4$.

In view of the great improvements in the Earth gravity field
modelling with the CHAMP \ct{pav} and, especially, GRACE
\ct{grace} missions an extensive search for alternative
combinations has been subsequently performed \ct{iorimp, iormor04,
iorMGM, iorproc, jas, ves}. In \ct{iormor04, iorMGM, iorproc} the
following combination has been proposed\footnote{The possibility
of using only the nodes of the LAGEOS satellites in view of the
improvements in the Earth gravity models from GRACE has been
propsed for the first time in \ct{grace}, although without
quantitative details. In \ct{ciucaz04} it seems that Ciufolini
refers to it as a proper own result with his reference [6] which
includes \ct{science98} of the present work and an announced
paper; \ct{science98} is not concerned with \rfr{iorform} because
it deals with \rfr{ciufform} and its analysis by means of EGM96.
Moreover, Iorio retains the e-mails in which he passed to
Ciufolini, with whom he was long in contact, the combination of
\rfr{iorform} along with the estimates of the systematic error
obtained with EIGEN2 and is disposed to make them publicly
available on request.} \eqi\delta\dot\Omega^{\rm LAGEOS }_{\rm obs
}+k_1\delta\dot\Omega^{\rm LAGEOS\ II}_{\rm obs}\sim \mu_{\rm
LT}48.2,\lb{iorform}\eqf where $k_1= 0.546$ and 48.2 is the slope,
in mas yr$^{-1}$, of the expected gravitomagnetic linear trend. It
has been adopted for the tests performed in \ct{ciucaz04} with the
2nd generation CHAMP-only EIGEN2 Earth gravity model \ct{eigen2}
and the 1st generation GRACE-only GGM01S \ct{ggm01} Earth gravity
model. \Rfr{iorform} allows to cancel out the first even zonal
harmonic $J_2$.
\section{The performed Lense-Thirring tests with the LAGEOS satellites}
The only performed tests aimed at the detection of the
Lense-Thirring precessions of \rfr{letprec} in the gravitational
field of the Earth with the existing LAGEOS satellites have been
performed, up to now, by Ciufolini and coworkers. They have used
the node-node-perigee combination of \rfr{ciufform} \ct{science98,
ciucaz04} and the node-node combination of \rfr{iorform}
\ct{ciucaz04}.

In \ct{ciucaz04} it is claimed that ``[...] {\it the
Lense-Thirring effect exists and its experimental value,}
[...]{\it , fully agrees with the prediction of general
relativity.}" in regard to both the tests with the EGM96 and
EIGEN-2 Earth gravity models. In this Section we will disprove
such statements.

The main objections to the results presented in these works can be
summarized as follows
\begin{itemize}
  \item Ciufolini has not performed tests by varying
  the length of the adopted observational time span, running
  backward and forward the initial epoch of the analysis, varying
  the secular rates of the even zonal harmonics in order to check
  their impact over different time spans, using different Earth
  gravity models in order to obtain a scatter plot of the obtained
  results.
  %Instead, the authors, for a given data set, have always used from time to time those Earth
  %gravity models which yielded just the closest results to what it is
  %a priori expected from the General Theory of Relativity.
  \item The total error budget has been
  underestimated, especially the systematic error of gravitational
  origin. E.g., the impact of the secular variations of the even zonal harmonics of the geopotential,
  which may become a very limiting factor over time spans many years long as those used, has not
  been
  addressed.
  Almost always 1$-\sigma$ results have been presented without any
  explicit indication of this fact.
  %\item All the relevant works of other authors, in which many of these issues have been addressed, have always been
  %consistently ignored
  %\item The node-node combination of \rfr{iorform} has been explicitly presented as a proper own result of Ciufolini
  %with
  %reference [6] of \ct{ciucaz04}, i.e. \ct{science98} of the
  %present paper,
  %which, instead, has nothing to do with \rfr{iorform}.
 \end{itemize}

\subsection{The node-node-perigee tests}
The combination of \rfr{ciufform} has been analyzed by using the
EGM96 \ct{egm96} Earth gravity model over 4 years in
\ct{science98} and over 7.3 years in \ct{ciucaz04}. The claimed
total error budget amounts to 20-25$\%$ over 4 years and to 20$\%$
over 7.3 years.
\subsubsection{The gravitational error}
The impact of the remaining uncancelled even zonal harmonics of
the geopotential $J_6, J_8, J_{10},...$ on \rfr{ciufform} has been
estimated by Ciufolini and coworkers with the full covariance
matrix of EGM96 in a root-sum-square calculation. In
\ct{science98} and, six years later, in \ct{ciucaz04} it is
claimed to be $\lesssim 13\%$. Apart from the fact that this is a
$1-\sigma$ level estimate, in \ct{ries}, as later acknowledged in
a number of papers \ct{iorimp, iorcelmec03, iorlar, iormor04,
iorproc, ves}, the use of the full covariance matrix of EGM96 has
been questioned. Indeed, it has been noted that in the EGM96
solution the recovered even zonal harmonics are strongly
reciprocally correlated; it seems, e.g., that the 13$\%$ value for
the systematic error due to geopotential is due to a lucky
correlation between $J_6$ and $J_8$ which are not cancelled by
\rfr{ciufform}. The point is that, according to \ct{ries}, nothing
would assure that the covariance matrix of EGM96, which is based
on a multi--year average that spans the 1970, 1980 and early 1990
decades, would reflect the true correlations between the even
zonal harmonics during the particular time intervals of a few
years adopted in the analyses by Ciufolini and coworkers. Then, a
more conservative, although pessimistic, approach would be to
consider the sum of the absolute values of the errors due to the
single even zonal as representative of the systematic error
induced by our uncertainty in the terrestrial gravitational field
according to EGM96 \ct{iorproc, iormor04}. In this case we would
get a conservative upper bound of 83$\%$ (1-$\sigma$). If a
root-sum-square calculation is performed by neglecting the
correlations between the even zonals a 45$\%$ 1-$\sigma$ error is
obtained \ct{iorcelmec03, iorproc, iormor04, ves}.

%These considerations are sufficient to rule out the Ciufolini's
%claims, still presented in \ct{ciucaz04}, about the combination of
%\rfr{ciufform}.
\subsubsection{The non-gravitational error}
Another important class of systematic errors is given by the
non--gravitational perturbations which affect especially the
perigee of LAGEOS II. The main problem is that it turned out that
their interaction with the structure of LAGEOS II changes in time
due to unpredictable modifications in the physical properties of
the LAGEOS II surface (orbital perturbations of radiative origin,
e.g. the solar radiation pressure and the Earth albedo) and in the
evolution of the spin dynamics of LAGEOS II (orbital perturbations
of thermal origin induced by the interaction of the
electromagnetic radiation of solar and terrestrial origin with the
physical structure of the satellites, in particular with their
corner--cube retroreflectors). Moreover, such tiny but insidious
effects were not entirely modelled in the GEODYN II software at
the time of the analysis of \cite{science98, ciucaz04}, so that it
is not easy to correctly and reliably assess their impact on the
total error budget of the measurement performed during that
particular time span. According to the evaluations in
\cite{luc02}, the systematic error due to the non--gravitational
perturbations over a time span of 7 years amounts to almost
28$\%$. However, according to \cite{ries}, their impact on the
measurement of the Lense--Thirring effect with the nodes of LAGEOS
and LAGEOS II and the perigee of LAGEOS II is, in general, quite
difficult to be reliably assessed.
%Moreover, also
%the formal, standard statistical error can be notably influenced
%if, e.g., the effect of the direct solar radiation pressure
%effects on the perigee of LAGEOS II is not suitably accounted for
%in the data analysis.

So, by adding quadratically the gravitational and
non--gravitational errors of \cite{luc02} we obtain for the
systematic uncertainty $\delta\mu_{\rm LT}^{\rm systematic}\sim
54\%$ if we assume a 45$\%$ error due to geopotential. The sum of
the absolute values of the errors due to gepotential added
quadratically with the non--gravitational perturbations would
yield a total systematic error of $\delta\mu_{\rm LT}^{\rm
systematic}\sim$ 88$\%$. It must be noted that the latter estimate
is rather similar to those released in \cite{ries}. Note also that
they are 1-$\sigma$ evaluations. Moreover, it should be considered
that the perigee of LAGEOS II is also sensitive to the eclipses
effect on certain non--gravitational perturbations. Such features
are, generally, not accounted for in all such estimates. An
attempt can be found in \cite{ves99} in which the impact of the
eclipses on the effect of the direct solar radiation pressure on
the LAGEOS--LAGEOS II Lense--Thirring measurement has been
evaluated: it should amount to almost 10$\%$ over an observational
time span of 4 years.
\subsection{The node-node tests}
In this Section we will deal with the node-node combination of
\rfr{iorform}. Such observable only cancels out the gravitational
bias of the first even zonal harmonic $J_2$, but has the great
advantage of discarding the perigee of LAGEOS II and its insidious
non-gravitational perturbations.

%It has explicitly been proposed for the first time in
%\ct{iormor04, iorMGM, iorproc}, posted on the WEB in April, August
%and September 2003, respectively. The possibility of using a
%LAGEOS-LAGEOS II node-only observable was presented for the first
%time in \ct{grace} without quantitative details, in the context of
%the expected improvements from the GRACE mission.
\subsubsection{The
gravitational error}\lb{zonrat} In \ct{ciucaz04} the node-node
combination of \rfr{iorform} has been analyzed with the 2nd
generation CHAMP-only EIGEN2 \ct{eigen2} and the 1st generation
GRACE-only GGM01S \ct{ggm01} Earth gravity models over a time span
of almost 10 years.

In \ct{iormor04} the impact of the static part of the
geopotential, according to the CHAMP-only EIGEN2 Earth gravity
model, is evaluated as $18\%$ (1-$\sigma$ root-sum-square
covariance calculation), $22\%$ (1-$\sigma$ root-sum-square
calculation) and $37\%$ (1-$\sigma$ upper bound). Ciufolini
reports 18$\%$ obtained in a root-sum-square fashion with the full
covariance matrix of EIGEN2 for which the same remarks as for
EGM96 holds. Moreover, he does not consider the fact that EIGEN2
is only based on six months of data and that the released sigmas
of the even zonal harmonics of low degree, which are the most
relevant in this kind of analyses with the LAGEOS satellites, are
rather optimistic, as explicitly pointed out in \ct{eigen2} and
acknowledged in \ct{iormor04}. In regard to the GGM01S model, the
covariance matrix was not publicly released. Ciufolini correctly
presents a 19$\%$ which is the 1$-\sigma$ upper bound obtained in
\ct{iormor04}. However, GGM01S is only based on 111 days of data.

In our opinion the author's conclusion ``We conclude, using the
Earth gravity model EIGEN-2S, that the Lense-Thirring effect
exists and its experimental value, $\mu=0.98\pm 0.18$, fully
agrees with the prediction of general relativity" is optimisitc.
Indeed, he claims that in his 18$\%$ total error budget all the
error sources are included. Ciufolini neglects the impact of the
time-dependent gravitational perturbations on \rfr{iorform}.
Indeed, they may turn out to be a serious limiting factor mainly
due to the secular variations of the even zonal
harmonics\footnote{The problem of the secular variations of the
even zonal harmonics in post-Newtonian tests of gravity with
LAGEOS satellites has been quantitatively addressed for the first
time in \ct{lucc03}. In regard to the
\leti\ measurement with \rfr{iorform}, it has been, perhaps,
misunderstood in \ct{iormor04}. } $\dot J_{\ell}$. Indeed,
\rfr{iorform} allows to cancel out $\dot J_2$, but is sensitive to
$\dot J_4$, $\dot J_6$,..., as pointed out in \ct{iorgrace}. The
uncertainties in the $\dot J_{\ell}$ are still quite large: see
Table 1 of \ct{cox}. From it the values of Table \ref{jdots} can
be inferred.

\begin{table}[Secular rates of the even zonal harmonics.
]\caption{Weighted means and standard deviations from Table 1 of
\ct{cox} of the secular rates of the first three even zonal
harmonics in units of $10^{-11}$ yr$^{-1}$.}\label{jdots}
\begin{tabular}{llll}
\noalign{\hrule height 1.5pt}

 & $\ell=2$ & $\ell=4$ & $\ell=6$  \\

\hline $\dot J_{\ell}$ & -2.113 & -0.6992 & -0.3594\\
$\sigma_{\dot J_{\ell}}$ & 0.0810 & 0.2029 & 0.1765\\

\noalign{\hrule height 1.5pt}
\end{tabular}
\end{table}
 On the other hand, their impact on the
\leti\ measurement grows linearly in time\footnote{For a possible
alternative combination which would cancel out the first three
even zonal harmonics along with their temporal variations see
\ct{jas, ves}.}. Indeed, the mismodelled shift, in mas, of
\rfr{iorform} due to the secular variations of the uncancelled
even  zonal harmonics can be written as \eqi
\sum_{\ell=4}\left(\dot\Omega_{.\ell}^{\rm LAGEOS}+k_1
\dot\Omega_{.\ell}^{\rm LAGEOS\ II }\right)\frac{\sigma_{\dot
J_{\ell}}}{2}T^2_{\rm obs},\lb{quadr}\eqf where the coefficients
$\dot\Omega_{.\ell}$ are $\partial \dot\Omega_{\rm class}/\partial
J_{\ell}$ and have explicitly been
 calculated up to degree $\ell=20$ in \ct{NC02, iorcelmec03}. It must
be divided by the gravitomagnetic shift, in mas, of \rfr{iorform}
over the same observational time span \eqi \left(\dot\Omega_{\rm
LT}^{\rm LAGEOS}+k_1 \dot\Omega_{\rm LT}^{\rm LAGEOS\ II} \right)
T_{\rm obs}=48.2\ {\rm mas\ yr^{-1}}\ T_{\rm obs}.\lb{dotmas}\eqf
 By assuming, e.g., $\sigma_{\dot J_4}=0.6\times 10^{-11}$
yr$^{-1}$ and $\sigma_{\dot J_6}=0.5\times 10^{-11}$ yr$^{-1}$
\cite{cox}, it turns out that the  percent error on the
combination \rfr{iorform} grows linearly with $T_{\rm obs}$ and
would amount to $1\%$ over one year at $1-\sigma$ level. This
means that, over 10 years, their impact  is $\sim 10\%$
(1-$\sigma$). In Section \ref{smerd} we will quantitatively
support this evaluation.
%Alternatively, if we look at
%the rate\footnote{Indeed, the normalized slope of the time series
%is measured. }, in mas yr$^{-1}$, these figures must be doubled.
%Indeed, the mismodelled secular rate due to the $\dot J_{\ell}$ is
%\eqi \sum_{\ell=2}\left(\dot\Omega_{.\ell}^{\rm LAGEOS}+k_1
%\dot\Omega_{.\ell}^{\rm LAGEOS\ II }\right)\delta\dot
%J_{\ell}T_{\rm obs},\lb{dota}\eqf which must be divided by the
%\leti\ secular trend \eqi \dot\Omega_{\rm LT}^{\rm LAGEOS}+k_1
%\dot\Omega_{\rm LT}^{\rm LAGEOS\ II} =48.2\ {\rm mas\
%yr^{-1}}.\eqf
%%Moreover, fitting and removing a parabolic noise superimposed to a
%%linear trend might alter the genuine relativistic linear signal
%%which we are interested in.
\subsubsection{The impact of the secular variations of the even zonal harmonics: a quantitative
estimate}\lb{smerd} Here we describe a numerical experiment aimed
at a quantitative evaluation of the impact of $\dot J_{\ell}$.

The first step consists in simulating the time series of
$\delta\Omega^{\rm LAGEOS}+k_1\delta\Omega^{\rm LAGEOS II}$ in
order to obtain the qualitative and quantitative features of
Figure 4 of \ct{ciucaz04}. It refers to EIGEN2 and shows the raw
residual time series with a straight line which fits it. The
post-fit residuals amounts to 12 mas. In our model, called Input
Model (IM), we include
\begin{itemize}
\item
LT$\equiv S_{\rm LT}t$ with $S_{\rm LT}=48.2$ mas yr$^{-1}$.
Lense-Thirring trend as predicted by the General Theory of
Relativity in order to simulate the fact that the residuals of
LAGEOS and LAGEOS II have been built up by dealing with the
gravitomagnetic force as a totally unmodelled feature
\item
ZONDOT$\equiv \sum_{\ell=4}^6\{r\}\left(\dot\Omega_{.\ell}^{\rm
LAGEOS}+c_1 \dot\Omega_{.\ell}^{\rm LAGEOS\ II} \right)\left(\rp{
\dot J_{\ell} }{2}\right)t^2$. Quadratic term due to the $\dot
J_{\ell}$ according to Table \ref{jdots}. The numbers $\{r\}$ are
randomly generated from a normal distribution with mean zero,
variance one and standard deviation one. Note that EIGEN2 does not
solve for $\dot J_{\ell}$. In \ct{ciucaz04} there are no details
about the values included in the dynamical force models of the
orbital processor; thus we treat the secular rates of the even
zonal harmonics as unmodelled features fully absorbed by the
residuals.
\item
ZONALS$\equiv p\left(\rp{x}{100}\right)S_{\rm LT}t$. Linear trend
with a slope of $x\%$ of the Lense-Thirring signal. For EIGEN2
$x=37$ (sum of the absolute values of the individual errors) is
assumed. The number $p$ is randomly generated as for the $\{r\}$
\item
TIDE$\equiv\sum\{a_c\}\sigma_{A_c}\cos\left[\left(\rp{2\pi}{P}\right)t+\{f_c\}\right]+
\sum\{a_s\}\sigma_{A_s}\sin\left[\left(\rp{2\pi}{P}\right)t+\{f_s\}\right]$.
Set of various tidal perturbations of known periods $P$. For the
impact of such kind of perturbations on the orbits of the LAGEOS
satellites see \ct{ior01}. The sets of numbers
$\{a_c\},\{a_s\},\{f_c\},\{f_s\}$ are randomly generated as $p$
and the $\{r\}$
\item
NOISE. White gaussian noise with variable amplitude which
simulates the observational errors of the laser-ranged measurement
\end{itemize}
The full IM used in our analysis is thus \eqi {\rm
IM=LT+ZONDOT+ZONALS+TIDE+NOISE.}\eqf
 We include in our model the
possibility of varying the length of the time series $T_{\rm obs
}$, the temporal step $\Delta t$ which simulates the orbital arc
length, the amplitude of the noise and of the mismodelling in the
perturbations and the initial phases of the sinusoidal terms in
order to simulate different initial conditions and uncertainties
in the dynamical force models of the orbital processors. More
precisely, the magnitude of the mismodelling in the various
effects is randomly varied within the currently accepted ranges
(1-$\sigma$) by using random numbers generated from a normal
distribution with mean zero, variance one and standard deviations
one. The same also holds for ZONALS because it is impossible to
know, a priori, the sign of the slope of the residual trend due to
the even zonal harmonics. The so built IM represents the basis of
our subsequent analyses.

\begin{figure}
\begin{center}
\includegraphics[width=13cm,height=11cm]{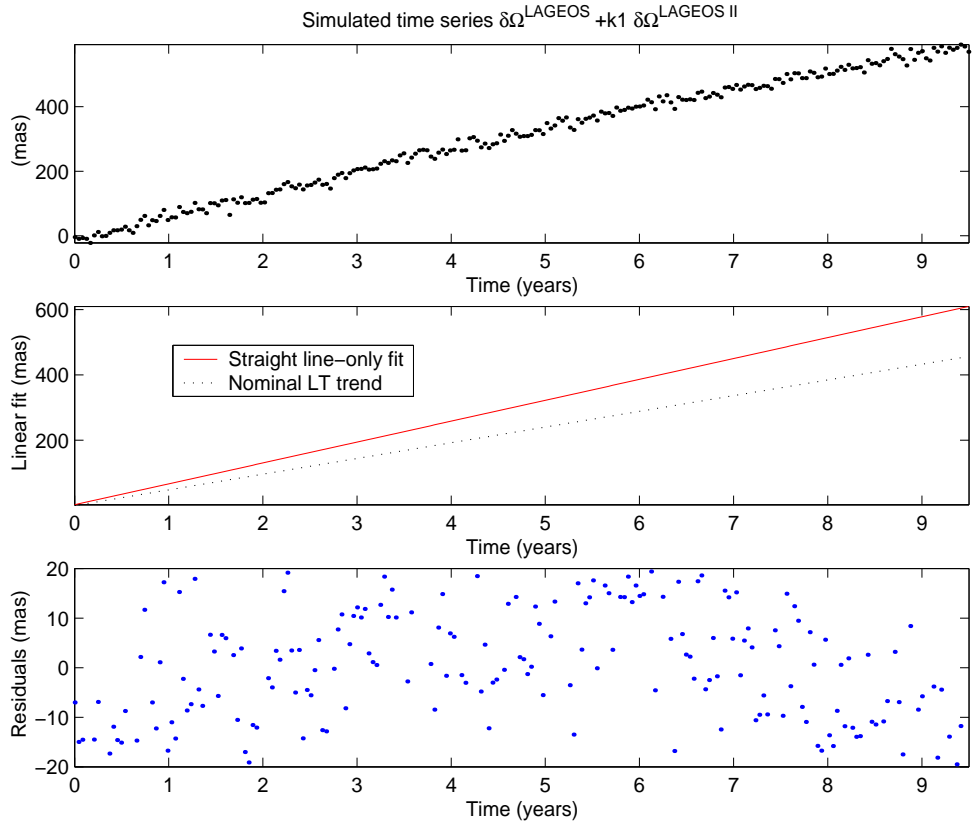}
\end{center}
\caption[Simulated time series, linear fit and post-fit residuals
over 9.5 years.]{\label{simul} Simulated time series, straight
line fit and post-fit residuals for $T_{\rm obs}=9.5$ years and
$\Delta t=15$ days. The RMS of the post-fit residuals is 12.7 mas.
The slope of the trend simulating the impact of the mismodelled
even zonal harmonics has been fixed to 37$\%$ of the
Lense-Thirring effect, according to EIGEN2. }
\end{figure}

With the so built IM we perform a set of 5000 runs by randomly
varying the initial phases, the noise and the mismodelling
amplitudes within the accepted intervals in order to simulate a
wide range of initial conditions and measurement errors which can
occur in the real world. The length of the time series is keep
fixed at $T_{\rm obs}=9.5$ years. In every run we fit the IM with
a straight line only (LF) and record the obtained slope $\mu$ with
the related error $\delta\mu$. Then, we calculate the averages of
$\mu$ and $\delta\mu$ and the standard deviation of the mean
$\Sigma$. For EIGEN2, i.e. $x=37$, we obtain
\eqi\left\langle\left|\rp{\mu-\mu_{\rm LT}}{\mu_{\rm LT
}}\right|\right\rangle\sim 30\%.\eqf Also the post-fit residuals
are calculated. Figure \ref{simul} shows the complete IM, its
straight line fit compared with the nominal Lense-Thirring trend
and the post-fit residuals for a given set of randomly chosen
initial conditions. The RMS post-fit amounts to 12.7 mas.
The averaged RMS post-fit is 11 mas. This shows that our procedure
represents a realistic starting point for our analyses.

A first interesting result is that the departure of the measured
slope $\mu$ from the nominal gravitomagnetic slope $\mu_{\rm LT}$,
which is included in IM, amounts to $\sim 30\%$ on average, while
for Ciufolini is $2\%$ only ($\mu=0.98$).

In order to evaluate the impact of the secular rates of the even
zonal harmonics in every run we also fit the IM with a quadratic
polynomial (QF) and compare the so obtained slope $\mu_{\rm QF}$
with the slope obtained in LF $\mu_{\rm LF}$. Note that procedure
is analogous to that adopted in \ct{pavior02} for the periodic
perturbations. On average, the difference between the two slopes,
i.e. the systematic error due to $\dot J_4,\dot J_6$, amounts to
\eqi\left<\left|\rp{\mu_{\rm LF}-\mu_{\rm QF}}{\mu_{\rm
LT}}\right|\right>\sim 14\%\eqf of the Lense-Thirring effect. This
result holds for $1-\sigma$. It is important to note that Figure
\ref{simul} has been obtained by assuming that the combined
residuals of the LAGEOS satellites absorb the quadratic signature
of $\dot J_{\ell}$ according to Table \ref{jdots} at 1-$\sigma$,
i.e. IM also includes ZONDOT. Nonetheless, it is difficult to
discern the parabolic signal which, instead, is present and  does
affect the recovery of the slope. This means that a simple visual
inspection of the plots of the combination of \rfr{iorform} cannot
be considered conclusive about the effect of $\dot J_4,\dot J_6$.

Another important point is that the combination of \rfr{iorform}
cannot be used to reliably constrain the zonals' rates by
measuring a $\dot J_4^{\rm eff}$. Indeed, it turns out that the
errors $\delta Q$ in the quadratic parameters of QF are always
larger than the estimated values themselves $Q$ and their mean
$<\delta Q/Q>$ over a given set of 5000 runs amounts to $\sim
260\%$ with a standard deviation of the mean of $68\%$. Moreover,
these figures change for different sets of 5000 runs.

If we repeat the same numerical experiments for $T_{\rm obs}=9.5$
years with $x=18$ (1-$\sigma$ root-mean-square full covariance
calculation) and $x=22$ (1-$\sigma$ root-mean-square variance
calculation) the situation does not substantially change
\eqi\left<\left|\rp{\mu_{\rm LF}-\mu_{\rm LT }}{\mu_{\rm LT
}}\right|\right>\sim 20\%,\ \left<\left|\rp{\mu_{\rm LF}-\mu_{\rm
QF }}{\mu_{\rm LT }}\right|\right>\sim 14\%.\eqf

Then, a more conservative 1-$\sigma$ estimate of the total
systematic error of the measurement of the Lense-Thirring effect
with the combination of \rfr{iorform} and the EIGEN2 Earth gravity
model is \eqi\delta\mu_{\rm LT}^{\rm total\ error}\leq 51\%.\eqf

While the forthcoming solutions from CHAMP and, especially, GRACE
will be able to improve the static part of the terrestrial
gravitational potential, i.e. the $J_{\ell}$, it is not so for
their secular rates $\dot J_{\ell}$. This fact sets for the
systematic error of gravitational origin a sort of threshold below
which it will not be possible to go unless much more accurate
determinations of $\dot J_4,\dot J_6$ will be available.

\section{Conclusions}
In this paper we have performed a detailed critical analysis of
the reliability and robustness of the so far performed tests aimed
to the detection of the Lense-Thirring effect in the gravitational
field of the Earth with the existing or proposed LAGEOS
satellites.

We can summarize our conclusions as follows
\begin{itemize}
\item
In regard to the node-node-perigee LAGEOS-LAGEOS II combination,
the claimed $20-25\%$ total accuracy obtained with the EGM96 Earth
gravity model, still presented in \ct{ciucaz04}, is not realistic
because of the impact of the non-gravitational perturbations on
the perigee of LAGEOS II and the mismodelling in the even zonal
harmonics of the geopotential whose $1-\sigma$ upper bound is
83$\%$.
\item In regard to the node-node LAGEOS-LAGEOS II
combination of \rfr{iorform}, extensive numerical tests have been
performed in order to quantitatively assess the impact of the
uncancelled secular variations of the even zonal harmonics on the
proposed measurement of the Lense-Thirring effect. A simulated
time series curve has been fitted with a straight line and a
quadratic polynomial and the so obtained slopes have been
compared. This procedure has been repeated over 5000 runs
performed by randomly varying the initial phases, the noise and
the mismodelling level within the currently accepted ranges of the
simulated signal. It turns out that the bias due to $\dot J_4$ and
$\dot J_6$ over 9.5 years amounts to $\sim 14\%$ on average. This
yields an upper bound of the total systematic error of the
performed tests with EIGEN2 of $\sim 51\%$.
\item Alternative combinations involving the use of existing
laser-ranged targets other than the LAGEOS satellites should be
analyzed. The most promising combination is, in principle, the one
that involves the nodes of LAGEOS, LAGEOS II, Ajisai and Jason-1
\ct{jas, ves}. It cancels out the first three even zonal harmonics
$J_2, J_4, J_6$ along with their temporal variations at the price
of introducing the relatively huge  non-gravitational
perturbations on Jason-1 which, however, should have a
time-dependent periodic signature with short periodicities.
According to the recently released combined
CHAMP+GRACE+terrestrial gravimetry/altimetry EIGEN-CG01C Earth
gravity model \ct{cg01c}, the systematic error due to the
remaining even zonal harmonics would amount to 0.7 $\%$
(root-sum-square calculation) and\footnote{The 1-$\sigma$ errors
for the node-node LAGEOS-LAGEOS II combination of \rfr{iorform}
are $5\%$ (root-sum-square calculation with the variance matrix)
and $6\%$ (upper bound), according to EIGEN-CG01C, while they are
$3\%$ and $4\%$, according to the 2nd generation GRACE-only
EIGEN-GRACE02S model.} 1.6$\%$ (upper bound) at 1$-\sigma$
\ct{jas}. The forthcoming more accurate and robust Earth gravity
model solutions from GRACE should especially improve the higher
degree even zonal harmonics, so that it might happen that the
difference between the node-only LAGEOS-LAGEOS II and the
node-only LAGEOS-LAGEOS II-Ajisai-Jason-1 combination will further
enforce to the detriment of the former one, at least in regard to
the gravitational error. The possibility of getting long time
series of the Jason's node should seriously be investigated with
real data tests. Moreover, Jason-1 is also affected by the orbital
maneuvers but they are mainly in-plane.
\end{itemize}

%\section*{Acknowledgments}
%I am grateful to J. Ries (Center for Space Research, CSR) for
%useful comments and discussions about the use of different Earth
%gravity models.
%-----------------------------------------

\end{document}